# OPTICS AND LASER PHYSICS

# Optical Spin Initialization of Nitrogen Vacancy Centers in a $^{28}$Si-Enriched 6H-SiC Crystal for Quantum Technologies

F. F. Murzakhanov[a,*] (ORCID: 0000-0001-7601-6314), M. A. Sadovnikova[a] (ORCID: 0000-0002-5255-9020),
G. V. Mamin[a] (ORCID: 0000-0002-7852-917X), D. V. Shurtakova[a] (ORCID: 0000-0003-4765-0724),
E. N. Mokhov[b] (ORCID: 0000-0002-2318-8849), O. P. Kazarova[b] (ORCID: 0000-0003-4077-4150),
and M. R. Gafurov[a] (ORCID: 0000-0002-2179-2823)

[a] *Institute of Physics, Kazan Federal University, Kazan, 420008 Russia*
[b] *Ioffe Institute, St. Petersburg, 194021 Russia*
*e-mail: murzakhanov.fadis@yandex.ru

High-spin defect centers in crystal matrices are used in quantum computing technologies, highly sensitive sensors, and single-photon sources. In this work, optically active nitrogen-vacancy color centers NV$^-$ in a $^{28}$Si-enriched (nuclear spin $I = 0$) 6H-$^{28}$SiC crystal have been studied using the photoinduced ($\lambda = 980$ nm) high-frequency (94 GHz, 3.4 T) pulsed electron paramagnetic resonance method at a temperature of $T = 150$ K. Three structurally nonequivalent types of NV$^-$ centers with axial symmetry have been identified and their spectroscopic parameters have been determined. Long spin–lattice, $T_1 = 1.3$ ms, and spin–spin, $T_2 = 59$ μs, ensemble relaxation times of NV$^-$ centers with extremely narrow (450 kHz) absorption lines allow highly selective excitation of resonant transitions between sublevels ($m_I$) caused by the weak hyperfine interaction ($A \approx 1$ MHz) with $^{14}$N ($I = 1$) nuclei for the quantum manipulation of the electron spin magnetization.

Intrinsic or artificially induced point defects in various materials are of particular interest because even extremely low their concentrations can significantly affect various physicochemical characteristics of materials. A number of vacancy defects (color centers) promote the variation of the visible and near-infrared optical properties of crystals [1–4]. The presence of a high-spin state (electron spin $S \geq 1$) in combination with optical, charge, and coherent properties makes it possible to consider color centers as candidates for a quantum bit (qubit), which a basic computing unit of a quantum computer. A nitrogen-vacancy center NV$^-$ in diamond is well known and studied and was used to implement prototypes of quantum computing devices and simulators, basic quantum cryptography protocols, and sensors sensitive to small changes in the temperature [1, 2], magnetic fields [3], and pressure [4]. However, the expensive production and low technological efficiency of a diamond matrix complicate extensive integration of diamond in the existing semiconductor devices. Silicon carbide SiC with color centers can be a possible alternative of diamond because it does not have the mentioned demerits and has sufficient mechanical, temperature, chemical, and radiation strengths [5]. It is known that SiC can be used as a reliable matrix for many high-spin ($S = 1$ and 3/2) defects such as a silicon vacancy V$_{Si}$, divacancy VV [6], and negatively charged nitrogen-vacancy centers NV$^-$, which are direct "twins" in microscopic structure of NV$^-$ centers in diamond [7, 8]. The luminescence spectrum of NV$^-$ centers in SiC with the wavelength $\lambda = 1.1$–$1.25$ μm (near infrared region) falls into the transparency window of biological tissues and optical fiber information transmission channels (O band), significantly extending the range of their application.

High-spin centers in molecular compounds and the crystals are characterized by the zero-field splitting parameters of spin sublevels $D$ and $E$ depending on the local symmetry, which are determined by spin–spin and spin–orbit contributions. The removal of spin degeneracy between the states with $M_S = 0$ and $M_S = \pm 1$ makes it possible to consider a color center as a qubit. Since color centers (NV$^-$ and VV) in SiC have a spin-dependent recombination channel (with spontaneous emission in the infrared region) in an optical excitation cycle, the state with $M_S = 0$ is predominantly populated. This effect is used to detect a signal from a single center by the optical detection of magnetic resonance and to develop highly sensitive nanosensors with a high spatial resolution [9].

In this work, we report the results obtained by photoinduced electron paramagnetic resonance (EPR) spectroscopy in the $W$ band (94 GHz) with a high spectral resolution and with the use of various pulse sequence to identify $NV^-$ centers, to determine the transverse and longitudinal electron magnetization relaxation times of $NV^-$ centers in the isotopically enriched 6H-$^{28}$SiC crystal.

Bulk 6H-$^{28}$SiC crystals enriched with the nonmagnetic $^{28}$Si isotope ($I = 0$) were grown by the high-temperature physical vapor deposition method using a precursor enriched with the $^{28}$Si isotope up to $\approx 99\%$ [10]. The concentration of the $^{29}$Si isotope ($I = 1/2$) in the grown samples estimated by the EPR method [11] is 1%, which is more than a factor of 4 lower than its natural abundance (4.67%). The concentration of the nitrogen impurity was $C \approx 10^{17}$ cm$^{-3}$. Then, the 6H-$^{28}$SiC crystals were irradiated by 2-MeV electrons with a fluence of $4 \times 10^{18}$ cm$^{-2}$ in order to form vacancy defects. The irradiated crystals were annealed in an argon atmosphere for 2 h at a temperature of $T = 900°C$ optimal to form stable $NV^-$ centers [12]. For the EPR study of $NV^-$ centers EPR, $450 \times 450 \times 670$-μm samples were cut from the crystals.

Pulsed EPR spectra were detected at a crystal temperature of $T = 150$ K on a Bruker Elexsys E680 commercial spectrometer (Karlsruhe, Germany) equipped with a helium flow cryostat and a superconducting magnet to generate a magnetic field on a sample up to $B_0 = 6$ T using the Hahn pulse sequence ($\pi/2-\tau-\pi$), where the $\pi/2$ microwave pulse had a duration of 40 ns and the time delay between the pulses was $\tau = 1.2$ μs. The integral intensity of the electron spin echo was measured with the variation of the magnetic field $B_0$. Transverse magnetization decay curves were obtained by detecting the electron spin echo at a $B_0$ value fixed at one of the spin transitions and by varying the time delay $\tau$. The longitudinal (spin–lattice) relaxation time was determined using the "inversion–recovery" sequence ($\pi - T_d - \pi/2 - \tau - \pi$) with the variation of the time $T_d$ between the inverting pulse and the detecting sequence. Photoinduced EPR experiments were carried out with a cw solid state laser with the wavelength $\lambda = 980$ nm and the output power up to 500 mW. The EPR spectra were simulated with the EasySpin 5.2.0 Matlab package [13].

In the absence of optical excitation (dark regime), the EPR signal was not detected (see the violet line in the lower inset of Fig. 1). The photoinduced EPR spectrum at the crystal orientation parallel to the external magnetic field ($c \parallel \mathbf{B}_0$) consists of two components separated by about 93 mT (see the green line in Fig. 1). The observed triplet shape of the spectrum is due to the splitting of spin levels in zero magnetic field characterized by the parameter $D$ (fine structure). The EPR spectrum includes several contributions caused by the existence of three different positions of carbon and silicon atoms in the crystal lattice of 6H-$^{28}$SiC, which differ in the local surrounding of the second coordination spheres. Thus, signals from three structurally nonequivalent $NV^-$ centers ($N_{k1}V_{k2}$, $N_hV_h$, and $N_{k2}V_{k1}$, see the green line in the lower inset of Fig. 1) with different spectroscopic characteristics summarized in Table 1 can be detected by the EPR method. The color centers were preliminarily identified using the results previously obtained in the $X$ band by stationary EPR methods [14]. The $D$ value is individual for each color center because of the sensitivity to the local environment and to the corresponding electron density distribution. The optical irradiation of the crystal at the wavelength $\lambda = 980$ nm results in the nonresonant excitation ($3A$, orbital singlet $\rightarrow 3E$, orbital doublet) of all three axial NV⁻ centers with the symmetry $C_{3v}$. A feature of the photoinduced EPR spectrum is the presence of both the absorption (low-field component) and emission (high-field component) signals; i.e., the signals are inverted in phase. This effect is due to the intercombination conversion at the transition of the center from the metastable excited orbital state ($1A$) to the ground state leading to the predominant population of the nonmagnetic state with $M_S = 0$ and, as a result, to the multiple enhancement of the EPR signal. The application of the strong magnetic field ($B_0 \approx 3.4$ T) such that the Zeeman interaction (94 GHz) dominates over splitting in zero magnetic field ($2D \approx 2.6$ GHz for the crystal orientation $c \parallel \mathbf{B}_0$), in contrast to optical detection of magnetic resonance or EPR in the classical $X$ band (9.4 GHz), promotes the formation of the degenerate triplet spin state with "pure" wavefunctions that are not mixed with the neighboring levels.

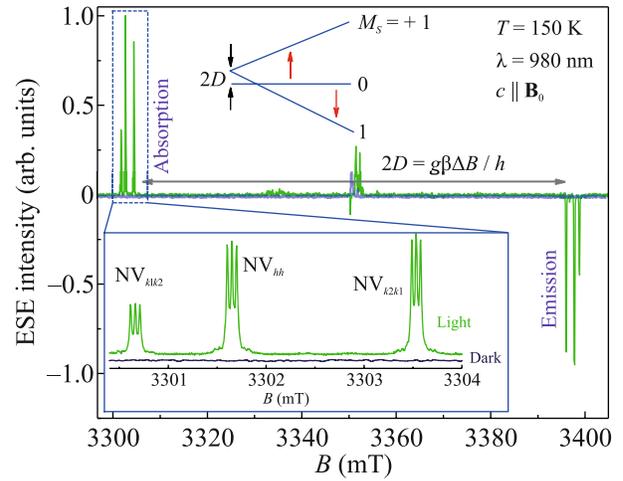

**Fig. 1.** (Color online) (Top panel) (Green line) Photoinduced EPR spectrum of $NV^-$ centers in the 6H-$^{28}$SiC crystal in the orientation $c \parallel \mathbf{B}_0$ and (violet line) the EPR spectrum in the dark regime in the absence of laser radiation. The inset shows the scheme of the energy levels of the color center with the variation of $B_0$, where $2D$ is the fine structure splitting. (Bottom panel) Detailed EPR spectrum of the low-field fine structure component with the attribution of line groups to three different nitrogen-vacancy centers.

**Table 1.** Spin Hamiltonian parameters for three axial $NV^-$ centers in 6H-$^{28}$SiC

| Defect | $g_\perp$ | $g_\parallel$ | $D$ (MHz) |
|---|---|---|---|
| $NV_{k1k2}$ | 2.0037 (3) | 2.0045 (3) | 1358 (2) |
| $NV_{hh}$ | 2.0035 (3) | 2.0045 (3) | 1331 (2) |
| $NV_{k2k1}$ | 2.0036 (3) | 2.0045 (3) | 1282 (2) |

Taking into account the symmetry of the point defect and the electron spin value, the following spin Hamiltonian was used to describe the results:

$$H = g\mu_B \mathbf{B} \cdot \mathbf{S} + D(S_z^2 - 2/3) + E(S_x^2 - S_y^2) \\ + A_\parallel S_z I_z + A_\perp(S_x I_x + S_y I_y) + P(I_z^2 - 2/3), \quad (1)$$

where $g$ is the spectroscopic splitting factor, $\mu_B$ is the Bohr magneton, $D$ and $E$ are the fine structure parameters, $S_{x,y,z}$ and $I_{x,y,z}$ are the projections of the electron and nuclear spins, respectively; and $A$ and $P$ are the hyperfine and quadrupole interaction constants.

The terms of the spin Hamiltonian (1) follow in the order of decreasing contribution and the last three terms refer to the electron–nuclear interaction, which is due to the presence of the magnetic $^{14}$N isotope with the nuclear spin $I = 1$ ($n = 2I + 1 = 3$). This interaction is manifested in the EPR spectrum in the form of an additional hyperfine structure with a splitting of about $A \approx 1$ MHz (see the lower inset of Fig. 1). The detailed study of the hyperfine and quadrupole interactions is beyond the scope of this work.

Extremely narrow EPR lines with a FWHM of only 450 kHz are particularly remarkable because they indicate a high quality of the crystal and a small spread in the $g$ and $D$ values. It is known that the parameter $D$ is very sensitive to local deformations [15], cluster structures [16], and other impurity centers [17] in the crystal lattice. The growth of the SiC crystal is also accompanied by a side effect of the formation of local crystal structures and polytype modifications differing from the main matrix, which also changes the parameter $D$ [18]. If the spin–spin contribution dominates in splitting in zero magnetic field, a small spread in the interatomic distance affects the spread of the parame-ter $D$ and finally leads to an additional inhomogeneous asymmetric broadening and to the distortion of structural lines [15]. The spread in the parameter $D$ was not detected in the studied crystal despite the defect nature of the color center with the predominant spin–spin splitting in zero magnetic field (which follows from the density functional calculations in [14, 19]).

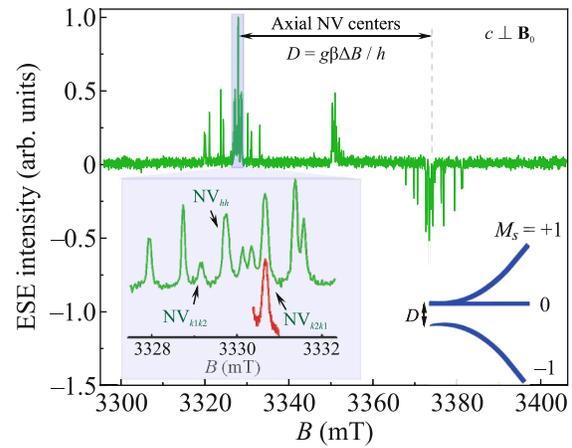

**Fig. 2.** (Color online) EPR spectrum in the orientation $c \perp \mathbf{B}_0$. The left inset shows the detailed spectrum of the low-field fine structure component (red line corresponds to the measurement with a smaller step in $B$). The right inset shows the scheme of the energy levels of the NV⁻ center in the orientation $c \perp \mathbf{B}_0$.

The rotation of the crystal from the parallel (0°, $c \parallel \mathbf{B}_0$) to perpendicular (90°, $c \perp \mathbf{B}_0$) canonical orientation with respect to $\mathbf{B}_0$ is accompanied by halving of the splitting in zero magnetic field (46.5 mT) (see Fig. 2). This corresponds to the angular dependence of the splitting of lines $\Delta B(\theta) = D(3\cos^2(\theta) - 1)$ for the axial symmetry of the defect ($C_{3v}$) without rhombohedral distortion ($E = 0$). The simulation of the position of resonant transitions for each NV⁻ center allowed us to determine the main spectroscopic parameters of the spin Hamiltonian summarized in Table 1. It is noteworthy that it is difficult to detect axial NV⁻ centers at the perpendicular orientation because of overlapping with absorption lines from basal NV⁻ centers with a lower symmetry ($C_{1h}$). As seen in the inset of Fig. 2, the hyperfine structure lines ($m_I$) at $c \perp \mathbf{B}_0$ are no longer resolved, forming a wide structureless line without the possibility of selective excitation.

Using standard pulse sequences, we determined the spin–spin and spin–lattice relaxation times at $T = 150$ K (see Fig. 3). This temperature was then established as optimal to observe photoactive defects with spin alignment under the optical excitation at 532 nm in the 4H-SiC and 6H-Si$^{13}$C crystal polytypes [6, 7]. Relaxation curves were recorded under the cw laser excitation with a low power of 125 mW to prevent the additional destruction of the coherence of NV$^-$ centers by the source of optical radiation. The transverse relaxation time of the magnetization of the NV$^-$ center is about 60 μs, which is longer than the value for the previously studied color centers (NV$^-$ $T_2 = 49$ μs and VV $T_2 = 40$ μs) for 4H-SiC even at lower liquid helium temperatures ($T = 7$ K) [7] and is four times longer than that for divacancies ($T_2 = 15$ μs) in the 6H-Si$^{13}$C crystal [6]. It is assumed that the main mechanism responsible for transverse relaxation is the spin–spin interaction between NV$^-$ centers. The $g$-factor for NV$^-$ centers is close to the value $g_e = 2.0023$ for the free electron, indicating that the spin–spin coupling makes an insignificant contribution and does not affect coherent properties [20]. According to Fig. 3b, the time $T_2$ for axial NV$^-$ centers in the perpendicular orientation decreases by a factor of more than 15 to $T_2 = 3.8$ μs. The relaxation (recovery) of the longitudinal magnetization of the color center is caused by the spin–lattice interaction and the corresponding time is $T_1 = 1.3$ and 0.2 ms for $c \parallel \mathbf{B}_0$ and $c \perp \mathbf{B}_0$, respectively. A significant difference between the relaxation times in two canonical orientations can be due to the spectral (spin) diffusion caused by the presence of resonant lines at $c \perp \mathbf{B}_0$ that overlap with the basal center lines. In the case of $T_2$, additional energy transfer (redistribution in the spin system for inhomogeneously broadened lines (see Fig. 2) occurs between structurally nonequivalent centers under the same, nearly resonant, conditions. The time $T_1$ is also affected by the spin diffusion through cross relaxation processes with transitions $\Delta M_S = \pm 1$ and $m_I = \pm 1$ appearing at the interaction between spin packets with small electron–nuclear coupling constants ($A, P \ll \nu_N$, where $\nu_N$ is the Larmor frequency of $^{14}$N nuclei). Since the longitudinal relaxation time is more than an order of magnitude longer than the transverse relaxation time, the spin–lattice coupling does not introduce additional loss of the phase coherence of the color center. Spin–lattice relaxation mechanisms (Orbach–Aminov, one- and two-photon Raman and "direct" processes) should be studied in detail with theoretical calculations and experiments in a wide temperature range. The analysis of the dynamic characteristics of the defect does not reveal the dependence of the relaxation times on the position of the NV$^-$ center within the error. For this reason, the results for only one NV$^-$ center with the most intense absorption spectrum are shown in Fig. 3. Since the EPR spectrum in the dark regime is absent, relaxation studies of NV$^-$ centers without cw optical excitation were not carried out. According to such measurements for 4H-SiC samples at $T = 7$ K [7], it can be expected that the detection of the electron spin echo will become possible at significantly lower temperatures.

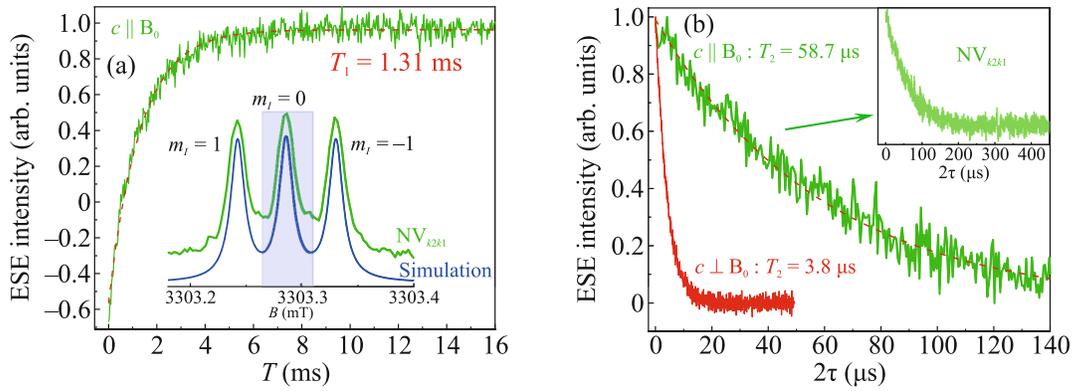

**Fig. 3.** (Color online) (a) Longitudinal magnetization recovery and (b) transverse magnetization decay curves of the NV$^-$ center in 6H-$^{28}$SiC and (red dashed lines) their single-exponential fits with the parameters indicated in the figure and main text. The inset in panel (a) shows the EPR spectrum of the NV$^-$ center at the $k_2k_1$ position and its simulation, where the rectangle marks the region of excitation and selective detection.

The longitudinal relaxation time of the NV$^-$ center allows experiments on the detection of signals by the electron–nuclear double resonance using radio-frequency and microwave pulses, e.g., in the Mims pulse sequence ($\pi_{MW}/2 - \pi_{RF} - \pi_{MW}/2 - \tau - \pi_{MW}/2 - \tau$ –electron spin echo) [7, 8]. In this case, to observe the stimulated electron echo, it will be possible to "separate" microwave pulses up to 1–1.5 ms without significant loss of the signal-to-noise ratio $S/N$. In this case, several radio-frequency pulses with a length up to 100 μs can be applied between these microwave pulses for the additional manipulation of nuclear spins and for the study of the electron–nuclear interactions.

To summarize, due to a high spectral resolution of the high-frequency EPR and to a decrease in the effect of the electron–nuclear interactions caused by mag-netic silicon nuclei, we have identified three structurally nonequivalent nitrogen-vacancy centers (electron spin $S = 1$) with axial symmetry ($C_{3v}$) and have determined the splitting in zero magnetic field $D$ for each defects. It has been shown that the nonresonant optical excitation ($3A \rightarrow 3E$) at $\lambda = 980$ nm leads to the effective spin alignment of the ground state of NV⁻ centers. Narrow EPR lines for the ensemble system allow the highly selective excitation of color centers with the formation of a "photon–spin" interface to fabricate elements of quantum spintronics. Long electron relaxation times of NV⁻ centers ($T_2 = 59$ μs and $T_1 = 1.3$ ms at $T = 150$ K) potentially make it possible to implement various quantum algorithms with the optical, microwave, and radio-frequency (multi)pulse interactions. The results obtained for the optical initialization and readout of states of NV⁻ centers using microwave pulse sequences open new possibilities of using high-spin photoactive centers in broadband semiconductor 6H-$^{28}$SiC crystals in quantum technologies.

## FUNDING

This work was supported by the Russian Science Foundation (project no. 24-22-00448).

## CONFLICT OF INTEREST

The authors of this work declare that they have no conflicts of interest.